\documentclass[aps, prb,
amsmath,amssymb,showpacs,groupedaddress]{revtex4}
\usepackage{graphicx}
\graphicspath{{../}}
\usepackage{epstopdf}
\begin{document}

\title{Flexoelectric deformation  of a homogeneously polarized ball}

\author{A. S.~Yurkov}
\affiliation{644076, Omsk, Russia, e-mail: fitec@mail.ru}
\date{\today}


\begin{abstract}

The problem of finding the elastic strains arising due to the converse flexoelectric effect is solved for a case of uniformly polarized ball of isotropic dielectric material. The strains occurs due to the fact that, as shown in a previous paper, in presence of flexoelectricity elastic boundary condition has  non-classical form.

\end{abstract}

\maketitle


\section{Introduction}
Direct flexoelectric effect is the linear response of the electric polarization to a gradient of elastic  strain. Although this effect has been  known for a long time (see \cite{bib:Tagantsev87} and references therein), some of its features have been understood only recently.  In particular,   according to general principles of thermodynamics  converse flexoelectric effect should also be present. At the same time, there is a certain asymmetry in the bulk constitutive electromechanical equations: while a linear strain obviously contributes to the homogeneous part of the polarization, homogeneous polarization does not appear in the equations that define the elastic stresses. Based on this, some authors have concluded that the flexoelectric effect, at least for a special arrangement, has no reversibility (see references in \cite{bib:Tag-Yur2011}).

The above paradox is resolved as follows \cite{bib:Tag-Yur2011,bib:Yurkov2011}. The equations of elastic equilibrium are differential, so they should be appended  by  boundary conditions. In the presence of flexoelectric effect elastic boundary conditions have non-classical form, and they include not only the gradient of polarization,
but also the polarization itself \cite {bib:Yurkov2011}.  This fact leads to the homogeneous polarization distorting the  body of finite size.

In the practical sense the importance of flexoelectricity is that it, in contrast to piezoelectricity,  is symmetrically allowed in centro-symmetric materials and therefore broadens the choice of materials which can be used for electromechanical devices. Besides, even  though flexoelectricity  is a weak  effect at the macroscale, at a nanoscale  it becomes  much stronger: reduced dimensions imply large gradients. This is why  flexoelectricity has recently  become a subject of intensive studies. Particularly flexoelectricity in nanowires and nanopills was  studied theoretically in \cite{bib:Eliseev}  but using classical elastic boundary conditions. 

Elastic boundary conditions obtained in \cite{bib:Yurkov2011}, were used in \cite{bib:MorozovskaAPL,bib:MorozovskaJAP}, but only for the flat part of the body surface. At the same time, it was shown that the curvature of the surface plays an important role in the boundary conditions \cite{bib:Yurkov2011}. It is  interesting  to solve the problem of flexoelectric deformation given a body surface curvature, which is the subject of this paper.  Here we restrict ourselves to the case of the spherical shape of the body of an isotropic dielectric. Although this is a very special case,  to solve the problem exactly for the more general case is difficult.  The exact solution, except that it is interesting in itself,  is useful as a test example in  development of approximate methods of solution for more general conditions.

Below, we also obtain  differential equations and boundary conditions in a manner different from \cite{bib:Yurkov2011},  using curvilinear coordinates. It is easier than converting the previously obtained boundary conditions to these   coordinates  that appear  most suitable for solving such problems.


\section{Equations of equilibrium and boundary conditions in curvilinear coordinates}

Here we need a mathematical tool which is almost identical to the one used in  general theory of relativity. The only difference is  in the fact that, since the space is flat, the Riemann tensor is equal to  zero, and the covariant derivatives commute. This mathematical tool  is described  in many standard textbooks,
here we make only a few brief comments to introduce symbols and definitions.

Curvilinear tensor indices are denoted by Greek letters. It will be convenient to assume that there is also a Cartesian coordinate system.  Cartesian indices are denoted by Latin letters. For curvilinear  indices is necessary to distinguish between  top and bottom, for the Cartesian indices there is no need for that as they will always be bottom. For repeated indices, unless otherwise noted, the Einstein summation convention  is adopted. The frame  vectors are defined as  $e_{i\alpha}= \partial r_i/\partial x^{\alpha} = r_{i,\alpha}\,$, where we use the notation of  partial derivative by the index separated by a comma. Here $ r_i $  are the Cartesian coordinates of the point and  $ x^{\alpha} \, $ are  the  curvilinear coordinates. Curvilinear tensor components are defined as $A_{\alpha\dots\beta}=A_{i \dots j}e_{i\alpha}\dots e_{j\beta}\, $. Raising of the Greek indices is performed in a standard way using the tensor $g^{\alpha\beta}\,$, which is reverse to the metric tensor $g_{\alpha\beta}=e_{i\alpha}e_{i\beta}\,$. Lowering of the Greek indices is performed by $g_{\alpha\beta}\,$. By $g$ we denote the determinant of $g_{\alpha\beta}\,$. Invariant volume element is given by  $d V = \sqrt{g} d^3 x \,$ , where $d^3 x=d x^1 d x^2 d x^3 \,$.

When equations in Cartesian components are transformed to  equations in curvilinear components, the usual partial derivatives are replaced by covariant derivatives, which are denoted by lower Greek indices  separated by semicolons.   Covariant derivatives are   defined   as
\begin{equation}
A_{\alpha_1\ldots \alpha_n ; \beta} = A_{\alpha_1\ldots \alpha_n , \beta} -
A_{\gamma\ldots \alpha_n }\Gamma^{\gamma}_{\alpha_1\beta} -  \,\,\ldots \,\, -
A_{\alpha_1\ldots \gamma}\Gamma^{\gamma}_{\alpha_n\beta} \, ,
\end{equation}
where
\begin{equation}
\Gamma^{\gamma}_{\alpha\beta} = e_i^{\gamma} e_{i\alpha,\beta}   =
\frac{1}{2} g^{\gamma\delta}
(g_{\delta\alpha,\beta}+g_{\delta\beta,\alpha}-g_{\alpha\beta,\delta}) \, 
\end{equation}
are Christoffel symbols.

Now we turn to the derivation of differential equations of equilibrium and boundary conditions.   The free energy is written in the form
\begin{equation}
 F= \int ( {\cal F}_{el}+{\cal F}_{p}+{\cal F}_{flx} ) \sqrt{g}d^3 x \, ,
 \end{equation}
where
\begin{equation}
{\cal F}_{el}  = \frac{1}{2}v^{\alpha\beta\gamma\delta\varepsilon\zeta}  u_{\alpha;\beta;\varepsilon}u_{\gamma;\delta;\zeta} +
\frac{1}{2} c^{\alpha\beta\gamma\delta}u_{\alpha;\beta}u_{\gamma;\delta} \,  ,
\end{equation}
\begin{equation}
{\cal F}_{p}= \frac{1}{2}a^{\alpha\beta}P_{\alpha}P_{\beta} +
\frac{1}{2}b^{\alpha\beta\gamma\delta}P_{\alpha;\beta}P_{\gamma;\delta} -
E^{\alpha}P_{\alpha} \, ,
\end{equation}
\begin{equation}
{\cal F}_{flx}=  \frac{1}{2}f^{\alpha\beta\gamma\delta } (P_{\alpha;\beta }u_{\gamma;\delta } - 
P_{\alpha} u_{\gamma;\delta;\beta })  \, . 
\end{equation}
The need for the higher terms in ${\cal F}_{el}$  which are proportional to $ u_{\alpha;\beta;\varepsilon}u_{\gamma;\delta;\zeta} $ was shown in \cite{bib:Yurkov2011}.

By varying the components of the polarization $ P_{\alpha} \, $ with vanishing boundary terms, we obtain differential equations of  polarization  equilibrium:
\begin{equation}
\label{p-eq}
a^{\alpha\beta} P_{\beta} -
b^{\alpha\beta\gamma\delta} P_{\gamma;\delta;\beta} - E^{\alpha} -
f^{\alpha\beta\gamma\delta }u_{\gamma;\delta;\beta } =  0 \, .
\end{equation}
Similarly, varying the elastic displacements, we obtain differential equations of elastic equilibrium:
\begin{equation}
\label{el-eq}
\sigma^{\alpha\beta}_{\phantom{\alpha\beta};\beta} = 0 \, ,
\end{equation}
where
\begin{equation}
\sigma^{\alpha\beta} = c^{\alpha\beta\gamma\delta}u_{\gamma;\delta} +
f^{\gamma\delta\alpha\beta } P_{\gamma;\delta } -
 v^{\alpha\beta\gamma\delta\varepsilon\zeta}u_{\gamma;\delta;\zeta;\varepsilon} \, .
\end{equation}

To derive the boundary conditions for these differential equations when the surface is free,  we require  the vanishing of additional boundary terms arising from the variation. Boundary conditions for the equations of polarization equilibrium  obtained  immediately:
\begin{equation}
\label{pbond}
\left(  b^{\alpha\beta\gamma\delta}P_{\gamma;\delta}n_{\beta}
+\frac{1}{2}f^{\alpha\beta\gamma\delta }u_{\gamma;\delta } n_{\beta} \right)_S = 0 \, ,
\end{equation}
where $n_{\beta}$ is unit vector normal to the surface.

As for the boundary conditions for the equations of elastic equilibrium,  an  integral equation  quite  similar to that  derived in \cite{bib:Yurkov2011} appear, and one needs to convert a surface integral. In \cite{bib:Yurkov2011}  such a  conversion was done using additional surface coordinate system. In this case, when the  curvilinear coordinates have been  introduced from the beginning, any additional surface coordinate system is not necessary. It is sufficient to require that the equation of the body surface has the form: $ x^3 = {\rm const} \,$. It is essential that the solution of specific problems  is convenient to do in  the coordinate systems of  just this type.

In this coordinate system, integration by parts is performed directly, and we obtain the elastic boundary conditions in the form:
\begin{equation}
\label{elbond1}
\Theta^{\alpha \beta \gamma}n_{\beta}n_{\gamma} = 0 \, .
\end{equation}
\begin{equation}
\label{elbond2}
\left( \sigma^{\alpha\gamma} - 
\Theta^{\alpha\beta\gamma}_{\phantom{\alpha\beta\gamma };\beta}
+\Theta^{\alpha\beta\delta}\Gamma^{\gamma}_{\delta\beta} + 
\left(\Theta^{\alpha 3\gamma}\sqrt{g}\right )_{,3}g^{-1/2}
 \right) n_{\gamma} = 0 \,  ,
\end{equation}
where
\begin{equation}
\Theta^{\alpha\beta\gamma}=
v^{\alpha\beta\varepsilon\delta\gamma\zeta}u_{\varepsilon;\delta;\zeta} -
\frac{1}{2}f^{\delta\gamma\alpha\beta } P_{\delta} \, .
\end{equation}
Naturally, these equations are only valid at the surface, we do not show it in the formulas for simplicity.

The boundary condition (\ref{elbond1}) is quite similar to that obtained in  \cite{bib:Yurkov2011}. Otherwise the condition (\ref{elbond2}) has a completely different look. Nevertheless, using (\ref{elbond1}) and performing a series of transformations, (\ref{elbond2}) can be reduced  to the form  similar to the second elastic boundary condition in \cite{bib:Yurkov2011}:
\begin{equation}
\label{elbond3}
\sigma^{\alpha\gamma}n_{\gamma}  - 
\Theta^{\alpha\beta\gamma}_{\phantom{\alpha\beta\gamma };\beta}n_{\gamma} +
\Theta^{\alpha\beta\gamma}_{\phantom{\alpha\beta\gamma };\delta }
n^{\delta}n_{\beta}n_{\gamma} 
-\Theta^{\alpha\beta\gamma}\gamma_{\beta\gamma} = 0 \, .
\end{equation}
 Here the tensor  $\gamma_{\beta\gamma}$ is expressed in terms of  Christoffel symbols:
\begin{equation}
\label{gamma-ab}
\gamma_{\beta\gamma} = \Gamma^{\varepsilon}_{\gamma\delta}n^{\delta}n_{\beta}n_{\varepsilon}
+\Gamma^{\varepsilon }_{\beta \delta} n^{\delta}n_{\varepsilon }n_{\gamma}
-\Gamma^{\delta}_{\gamma\beta}n_{\delta} 
 \,  .
\end{equation}
We emphasize that this relation is valid only in the coordinate system in which the equation of the surface is $ x^3 = {\rm const} \, $. The fact that this equality cannot be valid in an arbitrary coordinate system  is clear from the fact that the Christoffel symbols are not tensor components. Also it is possible to show  that the equation (\ref {gamma-ab}) is consistent with the equation for $ \gamma_ {\alpha \beta} $ derived in \cite{bib:Yurkov2011}. Moreover, one can show that the tensor $ \gamma_ {\alpha \beta} \, $ is symmetric, this  was not obvious from the equations written in \cite{bib:Yurkov2011}.

In general, equation (\ref{p-eq}) and (\ref{el-eq}) should be solved in the region occupied by the body with the boundary conditions (\ref{pbond}), (\ref{elbond1}) and (\ref{elbond3}). In order to determine the electric field $E_{\alpha}\,$,  the Poisson equation should be solved jointly,  within the body as  well as  outside. Solutions of the Poisson equation inside and outside the body should be  sewn together  with  the standard electrostatic boundary conditions. Electrostatic boundary conditions and Poisson equation in a curvilinear coordinate system are obvious, so we do not write them down.


\section{Homogeneously polarized ball of isotropic dielectric}

Now we will apply what was said in the previous section to the case of an isotropic dielectric ball. We use the conventional spherical coordinate system: $x=r\sin\theta\cos\psi \,$,  $y=r\sin\theta\sin\psi \, $, $z=r\cos\theta \,$.  Here $x\,$,  $y$ and $z$   are Cartesian coordinates,  $\psi=x^1\,$,   $\theta=x^2$ and $r=x^3$ -- curvilinear ones. Equation of the surface has the form $r = R\,$,  so that these curvilinear coordinates   belong to the  special class  described in the previous section. Instead of curvilinear  indices $1\, $, $2\, $ and $3\, $, we will use the indices $\psi\,$,  $\theta\,$  and $r\,$, respectively. So the  letters $r\, $,  $\psi$ and $\theta$ are excluded from notation of ``running'' indexes.

In Cartesian components  the second rank material tensor of an isotropic medium reduces to a scalar:  $a_{ij}=a\delta_{ij}\, $, where $\delta_{ij}\,$ is the Kronecker delta. Material tensors of the fourth rank are determined by two independent constants. So the elastic tensor can be written as $c_{ijkl}= c_{12}\delta_{ij} \delta_{kl} + c_{44} (\delta_{ik}\delta_{jl} + \delta_{il}\delta_{jk}) \, $,  the  tensors $b_{ijkl} $ and $f_{ijkl}\, $ have a similar form. Sixth rank tensor of  high elastic modulus of an isotropic medium is also determined by two independent constants:
\begin{equation}
\begin{array}{l}
\displaystyle
v_{ijlknm}  
=v_1(\delta_{ij}\delta_{lm}\delta_{nk} + \delta_{ij}\delta_{nl}\delta_{mk} +
\delta_{ij}\delta_{nm}\delta_{lk} 
 +\delta_{ik}\delta_{jn}\delta_{lm} + \delta_{il}\delta_{jn}\delta_{km} + 
\delta_{im}\delta_{jn}\delta_{kl}  + \\ 
\displaystyle
\quad\,\,\,   \quad\quad\quad\,\,\,
\phantom{+}\delta_{in}\delta_{jk}\delta_{lm} + \delta_{in}\delta_{jl}\delta_{km} +
\delta_{in}\delta_{jm}\delta_{kl} )  + \\
\displaystyle 
\quad\quad\quad\,\,\,
\phantom{+}\, v_2(\delta_{ik}\delta_{jl}\delta_{nm} + \delta_{ik}\delta_{jm}\delta_{nl}  
+\delta_{il}\delta_{jk}\delta_{nm} 
+ \delta_{il}\delta_{jm}\delta_{nk} + 
  \delta_{im}\delta_{jk}\delta_{nl} + \delta_{im}\delta_{jl}\delta_{nk} ) \, .
\end{array}
\end{equation}
To write these tensors in curvilinear components one need only to replace the Kronecker delta $\delta_ {ij}$ by the metric tensor with the corresponding location of the indices.

Using the general formulas for calculating the components of the metric tensor, its determinant and the Christoffel symbols, and also using the material tensors of an isotropic medium, one can obtain partial differential equations of the problem by  absolutely  straightforward  but cumbersome calculations. We will not write them down here because of their bulkiness and the fact that the calculations are quite standard. Instead, we will move to a discussion of  variables separation.

Generally the variable  separation  in such problems is performed by means of expansion of vectors in series in spherical vectors and expansion of scalars in series in scalar spherical harmonics. Note that, because of our  use of  covariant formalism here, the components  of spherical vectors do not coincide with the ones  usually used \cite{bib:Varshalovich}.  But the needed components can be easily  derived from the conventional  components.

In our case the expansion in spherical functions is essentially simplified: a series in spherical harmonics contains  only the terms with $ l = 1 \, $, $ m = 0 \, $.  Indeed, if far away from the ball the electric field is directed along the $ z $ axis and is homogeneous, then the electrostatic potential has only such a term. But then, since the variables are separated,   the other variables also have only such terms.  As a result, using the fact that $ Y_{10} (\psi, \theta) \sim \cos \theta \, $, it turns out that the angular dependence of the electrical potential is reduced to $ \cos \theta \, $;  the  $r$-components have the same dependence, the $\theta $-components are proportional to $\sin \theta \, $ and there are no $ \psi $-components. Thus, the problem is  reduced to  ordinary (radial) differential equations,   and it is only necessary  to provide a specified angular dependence explicitly.

The radial system of equations and the boundary conditions for the complete radial  problem (the ball in a uniform external electric field) is too cumbersome to write it here. Therefore, we restrict ourselves to a simplified version when the polarization is fixed and homogeneous. Thus only the elastic equations are solved. Since flexoelectric moduli of real materials are very small, this  is a good approximation.

In a simplified case, the system of radial differential equations has the form:
\begin{equation}
\label{eqf1}
\begin{array}{l}
\displaystyle
v_3(\xi^4 f''''_1+4\xi^3 f'''_1-8\xi^2 f''_1+16 f_1+ 8\xi^2 f''_2  
- 16f_2) +  \\
\displaystyle
\phantom{+} v_4(\xi^4 f''''_1+4\xi^3 f'''_1-6\xi^2 f''_1+12f_1 
 -2\xi^3 f'''_2+2\xi^2 f''_2+4\xi f'_2-12f_2) = \\
 \displaystyle
 \phantom{=} c_{44}\xi^2(2 \xi^2 f''_1+4\xi f'_1-6f_1 - 2 \xi f'_2+6 f_2) 
 + c_{12}\xi^2(\xi^2 f''_1+2\xi f'_1-2f_1 - 2\xi f'_2+2f_2)   \, ,
\end{array}
\end{equation}
\begin{equation}
\label{eqf2}
\begin{array}{l}
\displaystyle
v_3(\xi^4 f''''_2 + 4\xi^3 f'''_2 - 4\xi^2 f''_2 + 8f_2 + 
 4\xi^2f''_1 
 - 8f_1) -   v_4(2\xi^2 f''_2 - 4f_2 - \xi^3f'''_1 - 4\xi^2 f''_1 
 +2\xi f'_1 + 4f_1) = \\
 \displaystyle
\phantom{=}    c_{44}\xi^2(\xi^2 f''_2 + 2\xi f'_2  - 4f_2 
+\xi f'_1+ 4f_1) -   c_{12}\xi^2 (2f_2 - \xi f'_1 - 2f_1)  \, .
\end{array}
\end{equation}
Here $v_3=(v_1+2v_2)R^{-2}\,$, $v_4=(8v_1+4v_2)R^{-2}\, $, prime denotes the derivative with respect to the dimensionless radial coordinate  $\xi=r/R$, the components of the elastic displacement are  expressed in terms of $ f_1 (\xi) $ and $ f_2 (\xi) $ as follows: $ u_r = f_1 \cos \theta \, $, $ u_{\theta} =-R \xi f_2 \sin \theta \, $, $ u_{\psi} = 0 \, $. In the second equation we  introduced a factor $ r = R \xi $ so that $ f_2 (\xi) $ is an  analytic function at  zero for the final physical displacement in the center of the ball, as the $ g^{\theta \theta} = r^{- 2} \, $. The minus sign is added for convenience,  so that $ f_1 = f_2 = {\rm const} $ means a homogeneous vector field directed along the $z$-axis.

There are four boundary conditions for the   radial equations:
\begin{equation}
\label{bc1}
\begin{array}{l}
\displaystyle
 2v_1( f'''_2 +  f''_2 - 18 f'_2 + 34 f_2  +  8 f''_1 + 14 f'_1 
 - 34 f_1) +  2v_2(2 f'''_2 + 2 f''_2 - 20 f'_2 + 36 f_2  + \\ 
\displaystyle
\phantom{+} 4 f''_1 + 16 f'_1 - 36 f_1)  -   2c_{44}R^2 ( f'_2 - f_2 
+f_1) =   f_{12}R^2 P   \, , 
\end{array}
\end{equation}
\begin{equation}
\label{bc2}
\begin{array}{l}
\displaystyle
R^2 [c_{12}( 2 f_1 - 2 f_2)  + (c_{12} + 2c_{44})  f'_1  ]  -  v_1(9f'''_1 
+ 18 f''_1 - 46  f'_1 + 36 f_1   -   14  f''_2 + 32  f'_2 - 36 f_2) - \\ 
\phantom{-} 2 v_2(3f'''_1 + 6 f''_1 - 22  f'_1 + 28 f_1 -    2  f''_2  
+16  f'_2 - 28f_2)  = R^2 f_{12}P
  \, ,
\end{array}
\end{equation}
 \begin{equation}
 \label{bc3}
 (18 v_1 + 12 v_2)  f''_1 +
 12v_1 (3  f'_1 - 4 f_1 - 2  f'_2  
 + 4 f_2)=   (f_{12} + 2 f_{44}) R^2  P \, ,
 \end{equation}
 \begin{equation}
 \label{bc4}
 2 v_1( f''_2 + 2 f'_2 - 6f_2  + 2 f'_1 + 6f_1) +   4 v_2 ( f''_2   
 -2  f'_2 + 2 f_2 + 2  f'_1 - 2 f_1) =   f_{44}R^2  P \, .
 \end{equation}
 Note that the boundary conditions are non-uniform: the right hand side is proportional to the uniform polarization, which in this case is directed along the axis of $z$. The inhomogeneous boundary conditions  leads to a deformation of the body in this case.


\section{Solution of radial equations}

In general, for a system of two fourth-order equations (\ref {eqf1}) - (\ref {eqf2}) one needs  eight  boundary conditions, but here we have  only four  boundary conditions (\ref{bc1}) - (\ref{bc4}). However, as  usually in the case of such problems, the missing boundary conditions are replaced by the conditions that a solution should be analytic at zero. Therefore, the solution should be expressed in the form of series in non-negative powers of $\xi$.  However, it is more convenient  to find the  complete basis set of solutions, and then to present a general solution as a linear combination of such basis functions. If the  number of  basis functions  is equal to the number of boundary conditions,  then the coefficients in the linear combination are easily found by solving a system of linear algebraic equations.

In accordance with the above, we express a solution in the form:
\begin{equation}
f_i(\xi) = \sum_{k=1}^4 C_k{\cal B}_{ki}(\xi) \, ,
\end{equation}
\begin{equation}
{\cal B}_{ki}(\xi) = \sum_{n=0}^{\infty}a_{kin}\xi^n \, .
\end{equation}
The constant coefficients $a_{kin}$ are found from the condition that $ {\cal B}_{ki} (\xi) $ obeys the system of differential equations (\ref{eqf1}) - (\ref{eqf2}). This condition leads to the following linear algebraic equations for the coefficients (index $k$ which is numbering the solutions is omitted):
\begin{equation}
\label{aeq1}
\begin{array}{l}
\displaystyle
[v_3(n^4-2n^3-9n^2+10n+16)   
+v_4(n^4-2n^3-7n^2+8n+12)]a_{1\,n}  
+[v_3(8n^2-8n-16)  
-v_4(2n^3-\\ 
\displaystyle
\phantom{-} 8n^2+2n+12)] a_{2\, n}  
=[c_{44}(2n^2-6n-2) + c_{12}(n^2-3n)] a_{1 \, n-2} 
- [c_{44}(2n-10)+c_{12}(2n-6)]a_{2 \, n-2} \, ,
\end{array}
\end{equation}
\begin{equation}
\label{aeq2}
\begin{array}{l}
\displaystyle
[v_3(4n^2-4n-8)+ v_4(n^3+n^2-4n  
- 4)]a_{1\,n} + [v_3(n^4-2n^3-5n^2+6n+8) 
- v_4(2n^2-2n-4)]a_{2\,n} = \\
\displaystyle
\phantom{=}[c_{44}(n+2)  
+c_{12}n]a_{1\,n-2}+[c_{44}(n^2-3n-2) 
-c_{12}2]a_{2\,n-2} \, .
\end{array}
\end{equation}

From (\ref{aeq1}) - (\ref{aeq2}) one can see that the system of linear algebraic equations, even though it is infinite, has the characteristic  structure: the coefficients with greater  $n$ are expressed in terms of the coefficients with smaller $n$. Therefore one can find $a_ {in}$ for any finite $n \,$, one needs only to analyze the case of  small $n$, and define  some of the $a_ {in}$ for these small $n \, $. The analysis shows that if one  omits the physically meaningless solution corresponding to the translation of the ball as a whole along the axis  $ z $, then there are four independent solutions of (\ref{aeq1}) - (\ref{aeq2}) i.e. four basis functions  ${\cal B}_{ki} (\xi) \,$,  just the same number as the number of available boundary conditions. Note that one of the basis functions (say $ {\cal B} _ {1i} (\xi) \, $) can be chosen as a finite   series that contains only the second powers  of $ \xi \, $. Such a function obeys  not only the considered system of radial equations, but also the system of radial equations of the classical theory of elasticity. 

\begin{figure}[h]
\begin{center}
\includegraphics[width=12cm,keepaspectratio]{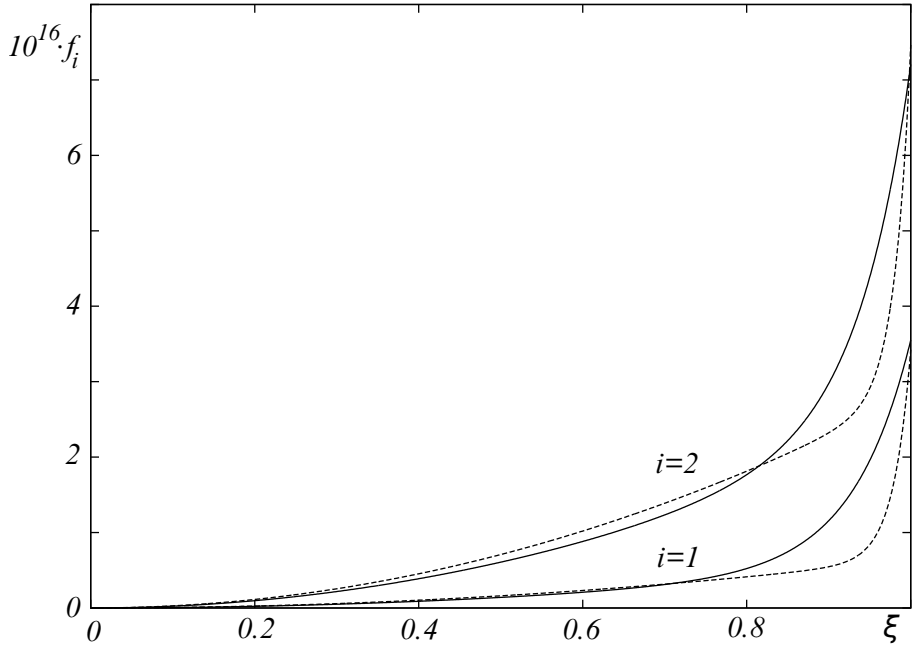}
\end{center}
%
\caption{Result of the calculation for $R=1\cdot 10^{-5}\,$, $P=1\,$,  $c_{44}=1.1\cdot 10^{12}\,$, $c_{12}=3.4\cdot 10^{12}\, $, $f_{44}=f_{12}=1\cdot 10^{-3}\, $. Solid line is for $v_1=2 \cdot 10^{-1}\,$, $v_2=1\cdot 10^{-1}\,$, dashed line is for $v_1=2\cdot 10^{-2}\,$, $v_2=1\cdot 10^{-2}\,$. }
\label{fig:figure}
\end{figure}

Without discussion of the following, rather conventional calculations, we give the numerical results (see the figure). Note that the graphs shows that, except for a  thin layer  near $\xi=1\, $, the curves are parabolic, the more so the  less  $v_1$ and $v_2\,$ are. In the bulk of the ball the only function  ${\cal B}_{1i}(\xi)\,$ remains with a good approximation.  Thus for a small $v_i$ except for a thin surface layer,  solution  approximately obeys the equations of classical elasticity theory. This observation is very worth to note because it tells how one can  built  an approximate method for solving such problems, which uses the classical theory of elasticity in the bulk of the body. Certainly, in the thin layer near the surface anyway one should use  equations of a  theory of elasticity  modified by flexoelectricity as described above. But a solution of this equations  for a thin layer is  simpler problem then the  solution of them for whole body. There is no  need for  such approach for  a ball, but it can be useful for more general problems.


\section{Conclusions}

In this paper we show how one can construct an exact, in the framework of the continuous medium theory, solution of the problem of flexoelectric deformation of homogeneously polarized ball of an isotropic dielectric.  This specific example  illustrates that the homogeneous polarization leads to flexoelectric deformation of the body, although the homogeneous part of the polarization does not enter in the differential equations of equilibrium. The reason for this deformation is related to non-classical form, in the presence of flexoelectricity, of the elastic boundary conditions, which  was previously shown in \cite{bib:Yurkov2011} in a general form.

Even for such a simple geometry the solution is  cumbersome and requires an introduction of non-standard special functions ${\cal B}_{ki}(\xi)\, $. However, the exact solution is useful, at least as a test example in the development of approximate methods and as a basis for the construction of a perturbation theory.


\end{document}